# Observation of thermal Hawking radiation at the Hawking temperature in an analogue black hole

Juan Ramón Muñoz de Nova, Katrine Golubkov, Victor I. Kolobov, and Jeff Steinhauer

*Department of Physics, Technion—Israel Institute of Technology, Technion City, Haifa 32000, Israel*

We measure the correlation spectrum of the Hawking radiation emitted by an analogue black hole and find it to be thermal at the Hawking temperature implied by the analogue surface gravity. The Hawking radiation is in the regime of linear dispersion, in analogy with a real black hole. Furthermore, the radiation inside of the black hole is seen to be composed of negative-energy partners only. This work confirms the prediction of Hawking's theory regarding the value of the Hawking temperature, as well as the thermality of the spectrum. The thermality of Hawking radiation is the root of the information paradox. The correlations between the Hawking and partner particles imply that the analogue black hole has no analogue firewall.

It was a profound realization that the entropy of a black hole [1] and Hawking radiation [2, 3] should have the same temperature, within a numerical factor on the order of unity. It was further asserted that Hawking radiation should have a thermal spectrum, which creates an information paradox [4, 5]. Furthermore, it was proposed that the physics of Hawking radiation could be verified in an analogue system [6]. This proposal was carefully studied and developed theoretically [7-19]. Classical white and black-hole analogues were also studied experimentally [20-23], as well as a variety of other analogue gravitational systems [24-30]. The theoretical works, combined with our long-term study of this subject [14, 31-34], allowed for the observation of spontaneous Hawking radiation in an analogue black hole [35]. Several theoretical works studied our observation [35], and made predictions about the thermality and Hawking temperature [36-40]. During the years since our observation [35], we have made many improvements to the experimental apparatus. This allows us to study the thermality of the Hawking spectrum, and compare its temperature with the prediction given by the analogue of the surface gravity. In this work, we find that the spectrum of Hawking radiation agrees well with a thermal spectrum, and its temperature is given by Hawking's prediction.

The analogue black hole consists of a flowing Bose-Einstein condensate. The flow velocity $v_{\text{out}}$ in the region $x < 0$ is less than the speed of sound $c_{\text{out}}$, as indicated in Fig. 1a. This region corresponds to the outside of the black hole. For $x > 0$, the flow is supersonic ($v_{\text{in}} > c_{\text{in}}$), corresponding the inside of the black hole. The sonic horizon, the point where $v = c$, is located at $x = 0$. In the inside region, the sound cones are tilted to the extent that all phonons travel inward, away from the sonic horizon.

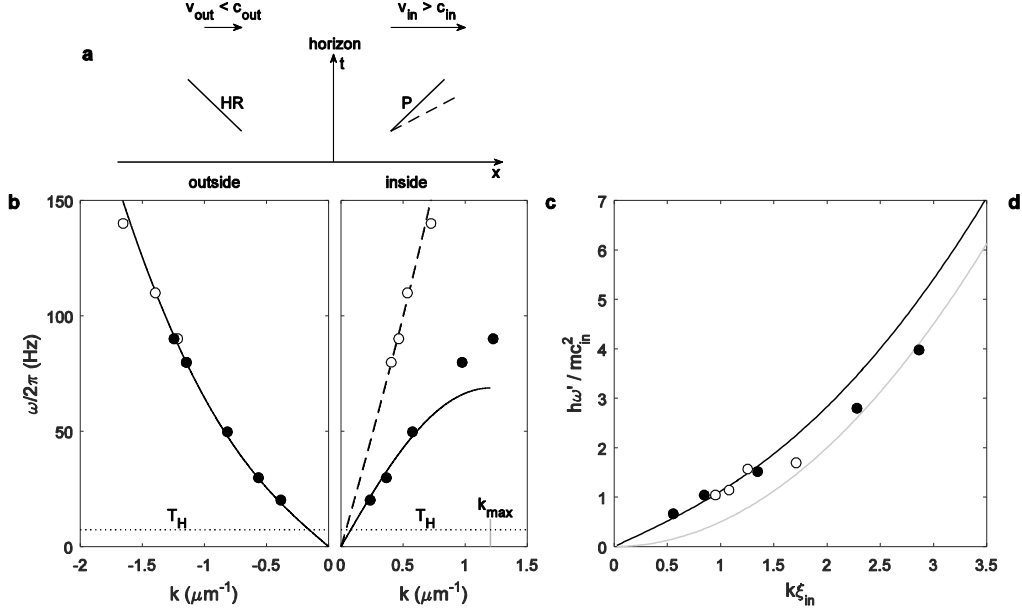

Figure 1. The Hawking and partner modes. a. A spacetime diagram showing the modes outgoing from the horizon. In addition to the Hawking mode and negative-energy partner mode indicated by solid lines, the dashed line indicates the positive-energy mode which is copropagating with the flow in the comoving frame. The partner and copropagating modes form a tilted sound cone inside of the analogue black hole. The horizontal arrows indicate the flow of the condensate. b-d. Dispersion relations. Circles are measurements from the oscillating horizon experiment. The solid and dashed curves are fits of Bogoliubov spectra including a Doppler shift. The dotted lines indicate the Hawking temperature. b. The dispersion relation outside the analogue black hole. c. The dispersion relation inside the analogue black hole. The filled circles and solid curve indicate the negative-energy partner mode. The open circles and dashed curve indicate the positive-energy copropagating mode. $k_{max}$ indicates the ultraviolet cutoff. d. The dispersion relation inside, in the "free-falling" comoving frame. The black curve is the Bogoliubov dispersion relation. The gray curve is the parabolic dispersion relation of a free particle.

For an analogue black hole, the Hawking temperature is given by $\hbar g/2\pi c$ [6], where the analogue surface gravity is $g = c(dv/dx - dc/dx)$ [8], and the derivatives and $c$ are evaluated at the sonic horizon. For a stationary one-dimensional flow, $nv$ is a constant, where $n$ is the one-dimensional density. We can thus write the Hawking temperature as

$$k_B T_H = -\frac{\hbar}{2\pi}\left(\frac{c}{n}\frac{dn}{dx} + \frac{dc}{dx}\right)\Big|_{x=0} \tag{1}$$

Eq. 1 is the predicted temperature of the Hawking radiation in an analogue black hole. It was derived using a linear dispersion relation, in analogy with massless particles emanating from a real black hole. In a Bose-Einstein condensate, the dispersion relation is linear in the low-energy limit. We should thus create an analogue black hole with sufficiently low Hawking temperature that the radiation is in the linear regime of the dispersion relation. We can then test whether the emitted Hawking radiation obeys the prediction (1). There are several theoretical works suggesting that this should be the case. Using the analytical results of Ref. 12 for a system similar to the experiment, we find that (1) gives an accurate prediction for $k_B T_H \lesssim 0.14\, mc_{out}^2$, where $m$ is the atomic mass. The experiment is within this limit since $k_B T_H = 0.12\, mc_{out}^2$. Ref. 36 studies our previous experiment [35], and concludes that the spectrum should be accurately Planckian, and the temperature should agree with the relativistic prediction of Eq. 1 within 10%. Ref. 37 also studies our previous experiment and finds that the temperature is expected to be quite close to Hawking's prediction (1).

We test Hawking's prediction (1) by measuring the spectrum of correlations between the Hawking and partner modes $\langle \hat{b}_H \hat{b}_P \rangle$, where $\hat{b}_H$ and $\hat{b}_P$ are the annihilation operators for Hawking and partner particles, respectively. Fortunately, $\langle \hat{b}_H \hat{b}_P \rangle$ is largely free of background. It represents correlations between the inside and outside of the black hole, and not many sources of such correlations come to mind, other than Hawking radiation. In contrast, any source of excitations can add to the background of the population $\langle \hat{b}_H^\dagger \hat{b}_H \rangle$. Indeed, the background of $\langle \hat{b}_H^\dagger \hat{b}_H \rangle$ is the difficulty in observing Hawking radiation from a real black hole. Since we work in the regime of low Hawking temperature and linear dispersion, we can use the $2 \times 2$ Bogoliubov transformation considered by Hawking, $\hat{b}_H = \alpha \hat{b}_+ + \beta \hat{b}_-^\dagger$ and $\hat{b}_P = \alpha \hat{b}_- + \beta \hat{b}_+^\dagger$, where $\hat{b}_+$ and $\hat{b}_-$ are annihilation operators for positive and negative-energy incoming modes, respectively, $|\alpha|^2 = |\beta|^2 + 1$, and $|\beta|^2 = 1/(e^{\hbar\omega/k_B T_H} - 1)$ [2, 3]. This immediately gives $\langle \hat{b}_H \hat{b}_P \rangle = \alpha\beta$ in the vacuum of incoming modes. Thus, we will compare our measurement of $|\langle \hat{b}_H \hat{b}_P \rangle|^2$ to $(|\beta|^2 + 1)|\beta|^2$ where $|\beta|^2$ is the Planck distribution at the predicted Hawking temperature (1).

The Hawking radiation is observed via the density-density correlation function $G^{(2)}(x, x') = \sqrt{\xi_{out}\xi_{in}/n_{out}n_{in}}\,\langle \delta n(x)\delta n(x')\rangle$ [9, 10]. We have found that $\langle \hat{b}_H \hat{b}_P \rangle$ is readily extracted from $G^{(2)}(x, x')$ by the relation [14]

$$S_0 \langle \hat{b}_H \hat{b}_P \rangle = \sqrt{\frac{\xi_{out}\xi_{in}}{L_{out}L_{in}}} \int dx\, dx'\, e^{ik_H x} e^{ik_P x'} G^{(2)}(x, x') \tag{2}$$

where $n_{\text{out(in)}}$ is the density outside(inside) the black hole, $L_{\text{out(in)}}$ is the length of the outside(inside) region, $\xi_{\text{out(in)}} = \hbar/mc_{\text{out(in)}}$, $x$ and $x'$ are in units of the healing length $\xi = \sqrt{\xi_{\text{out}}\xi_{\text{in}}}$, and $k_H$ and $k_P$ are the wavenumbers of the Hawking and partner modes respectively, in units of $\xi^{-1}$. The integral is performed over the region in the correlation function bounded by $-L_{\text{out}}/\xi < x < 0$ and $0 < x' < L_{\text{in}}/\xi$. The zero-temperature static structure factor is given by $S_0 = (U_{k_H} + V_{k_H})(U_{k_P} + V_{k_P})$, where $U_i$ and $V_i$ are the Bogoliubov coefficients for the phonons, which are not related to the Bogoliubov coefficients introduced by Hawking. It is natural that $S_0$ appears in (2) since the density amplitude of a phonon is proportional to $\sqrt{S_0}$ [32]. The only assumption in (2) is that modes with different frequencies are uncorrelated.

The analogue black hole is created in a Bose-Einstein condensate consisting of 8000 [87]Rb atoms. The elongated condensate is confined in a focused laser beam (red-detuned, $\lambda = 812$ nm, waist = 3.9 μm), propagating in the $x$-direction. We measure the transverse trapping frequency by applying a short pulse of a magnetic field gradient, which excites a monopole oscillation. We observe the oscillations as a function of time, giving a trapping frequency of 130 Hz at the focus of the laser. The region $x < 0$ is illuminated with an additional laser beam (blue-detuned, $\lambda = 442$ nm), which creates a positive potential, as illustrated in the inset to Fig. 2a. There is thus a downward potential step, a waterfall potential, near $x = 0$. The step potential moves at a constant speed, which is equivalent to the condensate flowing at a constant speed in the reference frame in which the step is stationary. The condensate flows over the step, which accelerates the $x > 0$ part of the condensate to supersonic speeds. The correspondingly lower density and speed of sound are seen in Figs. 2a and 2b. In this region, a phonon traveling toward the step in the "free-falling" frame (the frame comoving with the fluid) travels away from the step in the laboratory frame. The phonon is unable to reach the step, in analogy with a particle inside of a black hole. Thus, the region of the step is analogous to the event horizon.

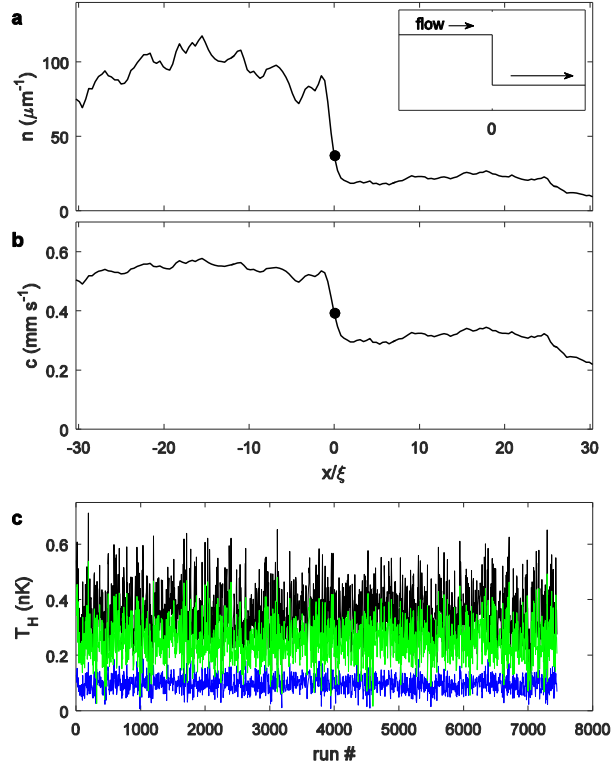

Figure 2. The profile of the analogue black hole. a. The density. The average of the 7400 runs of the experiment is shown. The circle indicates the horizon. The inset illustrates the step potential and the flow. b. The speed of sound by (3). c. The predicted Hawking temperature by (1). Each point is computed from the average of 5 runs of the experiment. The green and blue curves are the first and second terms in (1), respectively. The black curve is the predicted Hawking temperature, the sum of the two terms.

The experimental apparatus is that of Ref. 35, but with many improvements. One such improvement is a magnetic field environment with lower noise, as a result of improved power supplies which activate the 6 sets of magnetic field coils. This is important since static magnetic field gradients apply forces to the condensate via the Zeeman shift. Since the condensate is rather decompressed and weakly trapped in the axial direction, it is very sensitive to such gradients. In addition, 5 out of each 200 runs serve as reference images. For these images the step potential is not applied and the power of the focused laser trap is reduced, which further increases the sensitivity to an axial magnetic field gradient or to a tilt resulting in a gravitational gradient. The axial center of the reference images is found, and any axial shift is corrected by a slight fractional adjustment ($\leq 3 \times 10^{-4}$) of the current in one of the axial magnetic field coils. In addition, the optics has been improved, with reduced aberrations. This includes the optics for creating and translating the waterfall potential, as well as for imaging. Specifically, the system of lenses has been redesigned, and the acousto-optic modulator used to translate the waterfall potential has been replaced by a rotating mirror. Furthermore, there is improved mechanical stability in the magnetic field coils and optics. In addition, the temperature of the room has improved long-term stability.

The experiment is repeated 7400 times, giving a density profile $n(x)$ for each run. The ensemble-averaged $n(x)$ is shown in Fig. 2a. The circle indicates the location of the horizon. We find that the speed of sound $c(x)$ can be derived from $n(x)$ by the relation

$$c(x) = \sqrt{\frac{2\hbar a\omega_r(x)n(x)}{m}}\sqrt{\frac{1+3n(x)a/2}{(1+2n(x)a)^{3/2}} - \frac{\hbar\omega_{r_0}}{2U_0}} \qquad (3)$$

where $a$ is the scattering length, and $\omega_{r_0}$ and $U_0$ are the radial trapping frequency and potential depth respectively, at the laser focus. The first factor in (3) is the usual expression for a quasi-1D condensate with harmonic radial confinement. The second factor reduces the speed of sound due to the transverse degree of freedom. The first term in the square root in the second factor accounts for the finite density [41]. The second term accounts for the finite depth of the potential. By (3), we obtain the ensemble-averaged $c(x)$ shown in Fig. 2b. Furthermore, $n(x)$ and $c(x)$ are computed for the average of each 5 adjacent runs. The predicted Hawking temperature is then computed by (1), as shown in Fig. 2c, where each point on the curves corresponds to one set of 5 runs. The contributions of the first and second terms in (1) are indicated by green and blue curves, respectively. The total is indicated by the black curve. The average over the black curve gives a predicted Hawking temperature of 0.35 nK, or 0.12 $mc_{\text{out}}^2$.

We compare the density fluctuations at a pair of points $(x, x')$ in Fig. 2a over the ensemble of 7400 runs. Fig. 3a shows the resulting density-density correlation function for every pair of points. The dark band extending from the center of the figure is the correlations between the Hawking and partner particles. The angle of the band is 5.5° from the hydrodynamic angle. The latter is derived under the assumption that the correlations propagate at the wave speeds $c_{\text{out}} - v_{\text{out}}$ and $v_{\text{in}} - c_{\text{in}}$ outside and inside the black hole, respectively. The profile of the band in the $x''$ direction, averaged over the length of the band, is shown in Fig. 3b. Figs. 3c and 3d show a numerical simulation employing the truncated Wigner method [10], using parameters similar to the experiment. The results are seen to be similar to Figs. 3a and 3b. The maximum near $x'' = -2$ in Fig. 3b was also apparent in a previous numerical simulation [13]. The green curve in Fig. 3c shows the hydrodynamic path of Hawking and partner modes which travel at $c(x) - v(x)$ and $v(x) - c(x)$ outside and inside the black hole, respectively. The red curve shows the Hawking and copropagating modes, where the latter propagates with speed $v(x) + c(x)$. It is seen that the simulation is consistent with the Hawking/partner pairs (green curve), rather than the Hawking/copropagating pairs (red curve), as expected.

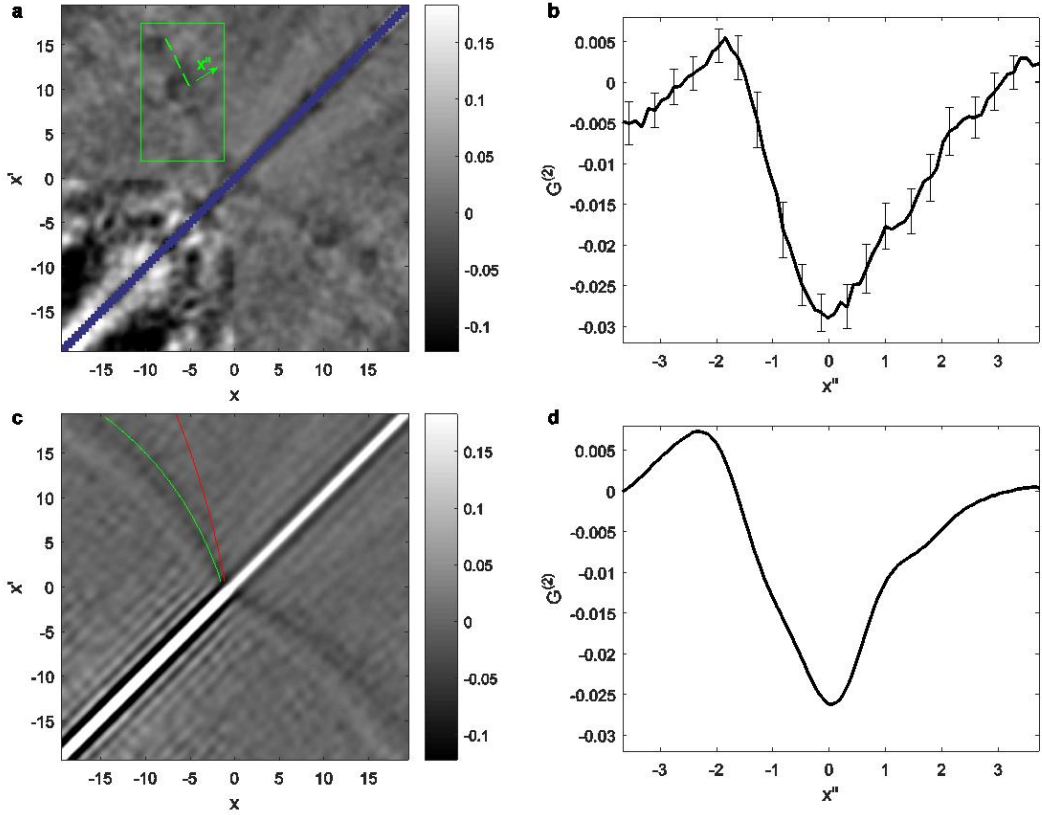

Figure 3. The measured Hawking radiation. a. The correlation function. The band extending from the origin is the correlations between the Hawking and partner particles. The green rectangle is the area used for the two-dimensional Fourier transform. The dashed line indicates the hydrodynamic angle for Hawking/partner pairs. b. The profile of the correlation band along the $x''$ direction in a. c and d. Numerical truncated Wigner simulations of a and b. The green curve in c corresponds to Hawking/partner pairs. The red curve corresponds to Hawking/copropagating pairs.

We extract the correlations between pairs of Hawking and partner modes by the Fourier transform (2). The Fourier transform is performed in the region outlined in green in Fig. 3a. The resulting correlation pattern $S_0|\langle \hat{b}_H \hat{b}_P \rangle|$ is shown in Fig. 4a. We obtain a one-dimensional plot of $S_0^2|\langle \hat{b}_H \hat{b}_P \rangle|^2$ from the Fourier transform of the profile in Fig. 3b by the relation $S_0 \langle \hat{b}_H \hat{b}_P \rangle = \sqrt{-\tan\theta - \cot\theta} \int dx'' e^{ikx''} G^{(2)}(x, x')$, where $\theta$ is the angle of the correlation feature in the $x$-$x'$ plane, measured relative to the positive $x$-axis [35]. The resulting correlation spectrum is shown in Fig. 4b.

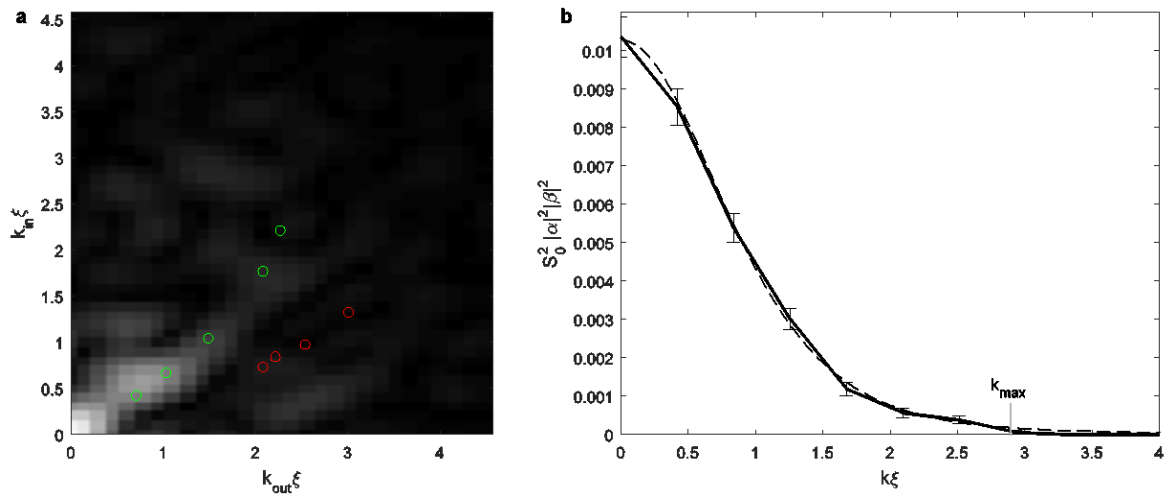

Figure 4. The spectrum of the Hawking radiation. a. The pattern of Hawking/partner correlations, computed within the green rectangle of Fig. 3a. The green circles indicate the Hawking/partner modes. The red circles indicate the Hawking/copropagating modes. The green and red circles are obtained from the driven oscillation experiment. b. The correlation spectrum of the Hawking radiation. The solid curve and error bars are the measured values. The dashed curve is the predicted thermal spectrum using the Hawking temperature from (1).

We have thus measured the predicted Hawking temperature, as well as the correlation spectrum of Hawking radiation as a function of wavenumber. In order to compare the two measurements, we need the relation between frequency and wavenumber, the dispersion relation. The dispersion relation is readily measured by the technique we introduced [35]. Waves are generated by causing the position of the step potential to oscillate with an amplitude of 0.5 µm. This oscillation at the horizon creates outgoing waves inside and outside the analogue black hole. The resulting dispersion relations are shown in Figs. 1b and 1c. The solid curves are fits of Bogoliubov dispersion relations including a Doppler shift, yielding $c_{out}$ = 0.52 mm sec$^{-1}$, $v_{out}$ = 0.23 mm sec$^{-1}$, $c_{in}$ = 0.31 mm sec$^{-1}$, and $v_{in}$ = 0.90 mm sec$^{-1}$. The fit misses the two highest points of the negative-energy (partner) branch of the dispersion relation in Fig. 1c. We can see the discrepancy more clearly by considering the frame which is comoving with the fluid inside the analogue black hole. In this frame, waves traveling to the left and right have the same dispersion relation, as seen in Fig. 1d. For larger $k$, it is seen that the measured points are not consistent with a spectrum of the Bogoliubov form. A qualitatively similar effect was observed [32] due to the modes in the radial direction [42, 43], but the effect of these modes is much smaller and occurs at much higher frequency than seen here. Rather, the outlying points in Fig. 1c are likely due to off-resonant stimulation of the excitations near the ultraviolet cutoff $k_{max}$.

Using the dispersion relation, we can express $|\beta|^2 = 1/(e^{\hbar\omega/k_B T_H} - 1)$ as a function of wavenumber. The resulting $S_0^2(|\beta|^2 + 1)|\beta|^2$ is indicated by the dashed curve of Fig. 4b. Very good agreement with the measured spectrum of Hawking radiation is seen, with no free parameters.

For a given applied frequency, the oscillating horizon experiment gives us $k_{\text{out}}$ corresponding to the Hawking mode, and the corresponding value or values of $k_{\text{in}}$ corresponding to the partner and/or copropagating modes, depending on the frequency. We indicate these measured pairs of modes ($k_{\text{out}}, k_{\text{in}}$) as circles in Fig. 4a. The green circles indicate the Hawking/partner modes, and the red circles indicate the Hawking/copropagating modes. The correlation pattern of the Hawking radiation (the gray region) is seen to lay along the green circles. The correlations are thus seen to be composed of Hawking/partner pairs, as expected. Since there are no correlations along the red circles, no correlations between Hawking and copropagating modes are seen. We can see that the experiment operates in the regime of linear dispersion since most of the correlations lay along the low-$k$, linear section of the green circles. This is also seen in Figs. 1b and 1c where the Hawking temperature indicated by the dotted line is in the linear section of the dispersion relations.

This work gives a quantitative confirmation of the temperature and thermality of Hawking radiation. These results were predicted in the literature for our system. The Hawking temperature is an important link between Hawking radiation and black-hole thermodynamics, since it also arises from considerations of entropy. The thermality of Hawking radiation suggests that one cannot obtain information about the inside of a black hole. This is the basis of the information paradox. The measurement is based on the correlations between the Hawking and partner modes. These correlations are seen to be of the predicted magnitude. Apparently, there is no analogue firewall at the horizon which would serve to reduce the correlations [44].

We thank the participants of the LITP Analogue Gravity Workshop for helpful conversations. We thank Iacopo Carusotto and Florent Michel for useful comments. This work was supported by the Israel Science Foundation.